\documentclass[12pt,a4paper]{article} 

\usepackage{amsmath,amssymb,cite,euscript}
\usepackage{epsfig,psfrag}

\def\pslash{\rlap{\hspace{0.02cm}/}{p}}
\def\vslash{\rlap{\hspace{0.02cm}/}{v}}

\newcommand{\MSbar}{\overline{\rm MS}}
\newcommand{\eps}{\epsilon}

\begin{document} 

\begin{titlepage}

\begin{flushright}
DESY 08-145\\ [0.2cm]
October 6, 2008 \\  
\end{flushright}

\vspace*{5mm}
\begin{center}
    {\baselineskip 25pt
    \Large\bf

\boldmath{NNLO corrections to $\bar B \to X_u  \ell \bar \nu $ in the
    shape-function region } 
    }

\vspace{0.8cm}
\begin{center}
{\sc H.~M.~Asatrian$^{(a)}$, C.~Greub$^{(b)}$,  and 
B.~D.~Pecjak$^{(c)}$}\\
\vspace{0.7cm}
$^{(a)}${\sl  Yerevan Physics Institute, 375036 Yerevan, Armenia } \\
\vspace{0.3cm}
$^{(b)}${\sl Center for Research and Education in Fundamental Physics,
    Institute for Theoretical Physics, Univ. Bern, CH-Bern, Switzerland   } \\
\vspace{0.3cm}
$^{(c)}${\sl Theory Group,
 Deutsches Elektronen-Synchrotron DESY,
 D-22603 Hamburg, Germany }\end{center}

\end{center}

\bigskip

\centerline{\bf Abstract}

\medskip 

\noindent 

The inclusive decay $\bar B \to X_u  \ell \bar \nu $ is of much
interest because of its potential to constrain the CKM
element $|V_{ub}|$.  Experimental cuts required to suppress
charm background restrict measurements of 
this decay to the shape-function region, where the hadronic final
state carries a large energy but only a moderate invariant mass.
In this kinematic region, the differential decay distributions satisfy a
factorization formula of the form $H \cdot J\otimes S$, where
$S$ is the non-perturbative shape function, and the object $H\cdot J$
is a perturbatively calculable hard-scattering kernel.  In this
paper we present the calculation of the hard function $H$ at 
next-to-next-to-leading order (NNLO) in perturbation theory.
Combined with the known NNLO result for  the jet function $J$,
this completes the perturbative part of the NNLO  calculation 
for this process. 

\end{titlepage}

\newpage

\section{Introduction} 
\label{sec:Introduction}

The inclusive decay $\bar B \to X_u  \ell \bar \nu $ is of much
interest because of its potential to constrain the CKM
element $|V_{ub}|$.  Due to experimental cuts required to 
suppress charm background, measurements of this decay are 
available only in the shape-function region, where the hadronic
final state is collimated into a single jet carrying a large 
energy on the order of $m_b$, and a moderate invariant 
mass squared on the order of $m_b \Lambda_{\rm QCD}$.  Much theoretical
effort has been put into establishing a factorization formalism
which enables the calculation of differential decay rates in 
this kinematic region. Early work in QCD was based on 
diagrammatic approaches \cite{Korchemsky:1994jb, Akhoury:1995fp}, 
whereas more recent papers 
\cite{Bauer:2003pi, Bosch:2004th, Lange:2005yw}
are based on soft-collinear effective theory (SCET) 
\cite{Bauer:2000ew, Bauer:2000yr, Beneke:2002ph}.   
The main result of these works can be summarized in the following
factorization formula for an arbitrary differential decay rate:  
\begin{equation}
\label{eq:VagueFact}
d\Gamma \sim H\cdot J\otimes S \, ,
\end{equation}
where the symbol $\otimes$ denotes a convolution.
The perturbative information  
is contained in the hard function $H$, which is related to physics at the 
hard scale $m_b$, and the jet function $J$, which is related to physics at the
intermediate scale $m_b \Lambda_{\rm QCD}$. The object  
$S$ is a non-perturbative shape function describing
the internal soft dynamics of the $B$ 
meson\cite{Neubert:1993ch, Bigi:1993ex}.  
The factorization formula is valid up to corrections in 
$\Lambda_{\rm QCD}/m_b$, which have been studied in detail in 
\cite{Lee:2004ja, Bosch:2004cb, Beneke:2004in}.  
The hard and jet functions to next-to-leading order (NLO) in perturbation
theory have been known for some time \cite{Bauer:2003pi, Bosch:2004th}, 
and the jet function at next-to-next-to-leading order (NNLO)
was obtained in \cite{Becher:2006qw}.   

The main purpose of this paper is to complete the perturbative
part of the NNLO corrections to the factorization formula 
(\ref{eq:VagueFact}) by obtaining the hard function to this order.  
The organization is as follows.  In Section \ref{sec:SCET}, we briefly
outline how to obtain the hard function through a matching calculation in
SCET.  The task is to extract three Wilson coefficients
$C_i$, which arise from integrating out the hard scale $m_b$ by
matching the semi-leptonic $b\to u$
transition current from QCD onto SCET.  The discussion there 
makes clear that the principle technical challenge is to 
calculate the two-loop QCD corrections to the $b\to u$ current.  
This loop calculation is the subject of Section 
\ref{sec:TwoLoop}, where we explain our calculational
procedure and give explicit results in terms of a set
of harmonic polylogarithms.  The method relies on a 
reduction to master integrals through integration-by-parts 
relations, which are then solved using differential equations. 
In Section \ref{sec:Wilson}, we use our results to obtain 
the Wilson coefficients $C_i$ at NNLO; 
a  phenomenological analysis of partial decay rates and the impact
on the determination of $|V_{ub}|$ is in progress
and will be presented in future work. We conclude in 
Section \ref{sec:Conclusions}.

\section{The hard function in SCET} 
\label{sec:SCET}

The QCD effects in inclusive semi-leptonic $B$ decays are contained in the 
hadronic tensor $W^{\mu\nu}$, from which any differential decay distribution 
can be derived.  It is defined as the discontinuity of the forward matrix
element of the current correlator $T^{\mu\nu}$, which is the time-ordered
product of two semi-leptonic $b\to u$ currents, 
$J^\mu = \bar u \gamma^\mu (1-\gamma_5)b$:
\begin{equation}
\label{eq:W}
W^{\mu\nu}=
\frac{1}{\pi} \, { \rm Im} \, 
\frac{\langle \bar B(v)|T^{\mu\nu}|\bar B(v)\rangle}{2 M_B}, 
\qquad 
T^{\mu\nu}=i \int d^4 x \, 
e^{i q \cdot x}{\rm T}\big\{J^{\dagger\mu}(0)J^{\nu}(x)\big\} \, .
\end{equation}
Here $q$ is the momentum carried by the lepton pair  and $v$ 
is the velocity of the $B$ meson.  Using the SCET formalism it 
is possible to show that the hadronic tensor obeys the factorization
formula
\begin{equation}
\label{eq:Wfact}
W^{\mu\nu}=\sum_{i,j=1}^3 H_{ij}(\bar n \cdot  p){\rm tr}
\left(\bar\Gamma_j \frac{\pslash_-}{2}\Gamma^\nu_i\frac{1+\vslash}{2}\right)
J\otimes S \, .
\end{equation}
We have introduced the vector $p \equiv m_b v-q$,
which in the parton model is the momentum of the final-state 
jet into which the $b$ quark decays, as well as its light-cone
decomposition,
\begin{equation}
p^{\mu}= (n\cdot p) \frac{\bar n^\mu}{2} + p_\perp^\mu 
+ (\bar n \cdot p)\frac{n^\mu}{2}\equiv p_+^\mu+p_-^\mu+p_\perp^\mu \, ,
\end{equation}  
where $n$ and $\bar n$ are two light-like vectors satisfying 
$\bar n \cdot n =2$.  The object $H_{ij}$ is defined as  
\begin{equation}
\label{eq:Hmatrix}
H_{ij}(\bar n \cdot p) = C_i(\bar n \cdot p) C_j(\bar n \cdot p) \, ,
\end{equation} 
where the Wilson coefficients $C_i$ arise from matching the semi-leptonic
$b\to u$ current from QCD onto SCET.  In position space and to 
leading order in the heavy-quark limit, this matching
is of the form
\begin{equation}
\label{eq:PosMatching}
e^{-im_b v \cdot x} \bar u(x)\gamma^\mu (1-\gamma_5)b(x) =
\sum_{i=1}^3 \int ds \, \tilde C_i(s)\bar \chi(x+s\bar n)\Gamma^\mu_i {\cal H}(x_-),
\end{equation} 
where we have followed the SCET conventions of \cite{Bosch:2004th}.  
The $\Gamma^\mu_i$ are a set of three Dirac structures, 
which we shall choose as
\begin{equation}
\label{eq:SCETdirac}
\Gamma_1^\mu = \gamma^\mu(1-\gamma_5), \qquad \Gamma_2^\mu = v^\mu(1+\gamma_5),
\qquad \Gamma_3^\mu = \frac{n^\mu}{n\cdot v}(1+\gamma_5) \, .
\end{equation}
In practice, the matching calculation is carried out in momentum space and 
yields results for the Fourier-transformed coefficients, which read
\begin{equation}
C_i(\bar n \cdot p) = \int  ds \, e^{is\bar n\cdot p} \, \tilde C_i(s)\,.
\end{equation}

The matching coefficients are obtained by evaluating UV-renormalized 
matrix elements of both sides of (\ref{eq:PosMatching}), corresponding
to calculations in full QCD and SCET.  The calculation 
is simplest when the external states are chosen as on-shell quarks
and both UV and IR divergences are regulated in dimensional regularization
in $d=4-2\epsilon$ dimensions.  In that case the loop corrections
to the SCET matrix elements are given by scaleless integrals and vanish,
so that the result is just its tree-level value
multiplied by renormalization factors from operator and 
wave-function renormalization. The QCD result is written
in terms of three Dirac structures multiplied by scalar form factors, 
which we shall define according 
to
\begin{eqnarray}
\label{eq:DiDef}
\langle u(p)|J^\mu | b(p_b)\rangle &=& D_1 \bar u(p)\gamma^\mu (1-\gamma_5)u(p_b)
+ D_2 \bar u(p)\frac{p_b^\mu}{m_b} (1+\gamma_5) u(p_b) \nonumber \\
&+& D_3 \bar u(p) \frac{p^\mu}{m_b} (1+\gamma_5) u(p_b) \, ,
\end{eqnarray} 
where $u(p)$ and $u(p_b)$ are on-shell spinor wave functions, 
$p_b$ and $p$ are the momenta of the $b$ and $u$ quarks respectively,
and $p^2=0$, $p_b^2=m_b^2$.  We shall always work in the reference frame
where the perpendicular components of the external momenta vanish, 
and where $p_b^\mu = m_b v^\mu$ and $p^\mu = (\bar n \cdot p)n^\mu/2$.
Then the three Dirac structures multiplying the $D_i$ correspond to those 
in (\ref{eq:SCETdirac}) in an obvious way.

To determine the Wilson coefficients $C_i$ we also need the SCET 
matrix element, for which we can make an important simplification.  In general,
the result involves a renormalization matrix $Z_{ij}$ applied
to the bare SCET current operators. However, we can use that the partonic 
expression for the quantity $J\otimes S$ in the factorization formula
(\ref{eq:Wfact}) for the hadronic tensor is independent of the
coefficients $H_{ij}$ that multiply it.  This implies that 
the operator renormalization matrix is just the unit matrix 
multiplied by a single scalar factor $Z_J$.  Moreover,
for on-shell matching the wave function renormalization factors in SCET
are unity, and the SCET spinor wave functions correspond to those in QCD.    
Therefore, the coefficients $C_i$ can be obtained through the relations
\begin{eqnarray}
\label{eq:Matching}
C_i(\bar n \cdot p) &=&   \lim_{\epsilon\to 0}\,Z_J^{-1}(\epsilon,m_b, \bar n \cdot p, \mu)
D_i(\epsilon,m_b, \bar n \cdot p, \mu) \qquad (i=1,2) \, , \nonumber \\
C_3(\bar n \cdot p) &=&   \lim_{\epsilon\to 0}\,Z_J^{-1}(\epsilon,m_b, \bar n \cdot p, \mu)
\frac{p_b\cdot p}{m_b^2} D_3(\epsilon,m_b, \bar n \cdot p, \mu) \, .
\end{eqnarray}
The renormalization factor $Z_J$ can be determined in two different ways. 
The first is to require that the 
matching relation (\ref{eq:Matching}) is free of IR poles in dimensional 
regularization, which allows one to deduce the UV structure of the 
SCET currents from the IR structure of the $D_i$.   
A second method is to determine the UV poles of the 
object $J\otimes S$ in the parton model, using the two-loop 
anomalous dimensions for the  jet and soft functions, calculated 
in \cite{Becher:2006qw} and \cite{Grozin:1994ni, Becher:2005pd}.  
Agreement between the two methods is an important check on 
the factorization formalism, and also on the 
two-loop calculation of each function. The agreement will be 
verified in Section \ref{sec:Wilson} below.

We end this section by pointing out a subtlety in the matching 
calculation related to heavy-quark loops, 
which first becomes relevant at NNLO.  
Whereas the partonic matrix elements in QCD are calculated as an
expansion in $\alpha_s$ in the
$\MSbar$ renormalization scheme in a five-flavor theory,
where $n_f=n_l+n_h$ with $n_h=1$ for the $b$ 
quark, in SCET $b$-quark loops are absent and the 
matrix elements are calculated as an expansion in a four-flavor theory. 
To match results in the two theories as in 
(\ref{eq:Matching}), it is necessary to express
the UV renormalized results in five-flavor QCD in terms
of the four-flavor parameters of SCET.  
To achieve this, one renormalizes the 
coupling constant in the $n_f=n_h+n_l$ flavor
theory according to $\alpha_s^{\rm bare}=Z_{\alpha}^{n_h+n_l}\alpha_s$, 
with~(see e.g. \cite{Steinhauser:2002rq})
\begin{equation}\label{eq:Znhnl}
Z_{\alpha}^{n_h+n_l}=1-\frac{\alpha_s}{4\pi \epsilon}
\left[\frac{11}{3}C_A-\frac{4}{3} T_R n_f + \frac{4}{3} T_R n_h(1-N_\epsilon)\right].
\end{equation}
The function $N_\epsilon$ is fixed such that 
$\alpha_s$ is the $\overline{\rm MS}$-renormalized coupling
in the {\it four} flavor theory. Its value is
\begin{equation}
\label{eq:Znfnl}
N(\epsilon)=e^{\gamma\, \eps}
\left(\frac{ \mu^2}{m_b^2}\right)^\epsilon \Gamma(1+\epsilon)\,.
\end{equation} 
Results for the scalar amplitudes $D_i$ 
in this renormalization scheme can be obtained from
those in the  $\MSbar$ scheme in five-flavor QCD by making the replacement 
\begin{equation}
\label{eq:decoupling}
\alpha_s \to 
\alpha_s\left(1+
\frac{ \alpha_s}{4\pi} \frac{8}{3} T_R n_h\left[L+
\epsilon\left(L^2 +\frac{\pi^2}{24}\right)
+\epsilon^2\left( \frac{2L^3}{3}+\frac{\pi^2}{12} L- 
\frac{\zeta_3}{6}\right)\right]\right)
+\dots \, ,
\end{equation}  
where $L=\ln\mu/m_b$.  After applying this decoupling to the $D_i$,
dependence on $n_h$ in the pole terms, and thus $Z_J$, drops out.
This must be the case, since heavy quark loops do not exist in SCET,
where the $b$ quark field is treated as in HQET.  This same procedure 
was used in the completely analogous case of matching the 
$b\to s$ current at $q^2=0$ in  \cite{Ali:2007sj}.

From the above discussion, it is obvious that the main technical 
obstacle to obtaining the Wilson coefficients $C_i$ is the calculation 
of the QCD form factors $D_i$.  
This will be the subject of the next section.

\section{Two-loop QCD corrections to the $b\to u$ current } 
\label{sec:TwoLoop}

In this section we perform the calculation of the 
renormalized scalar form factors $D_i$ at two-loop
order.  We begin by outlining the calculational procedure in 
Section \ref{subsec:QCDcalculation}, and then give the final
results in Section \ref{subsec:QCDresults}.

\subsection{Calculational procedure }
\label{subsec:QCDcalculation}

\begin{figure}[t]
\begin{center}
\epsfig{file=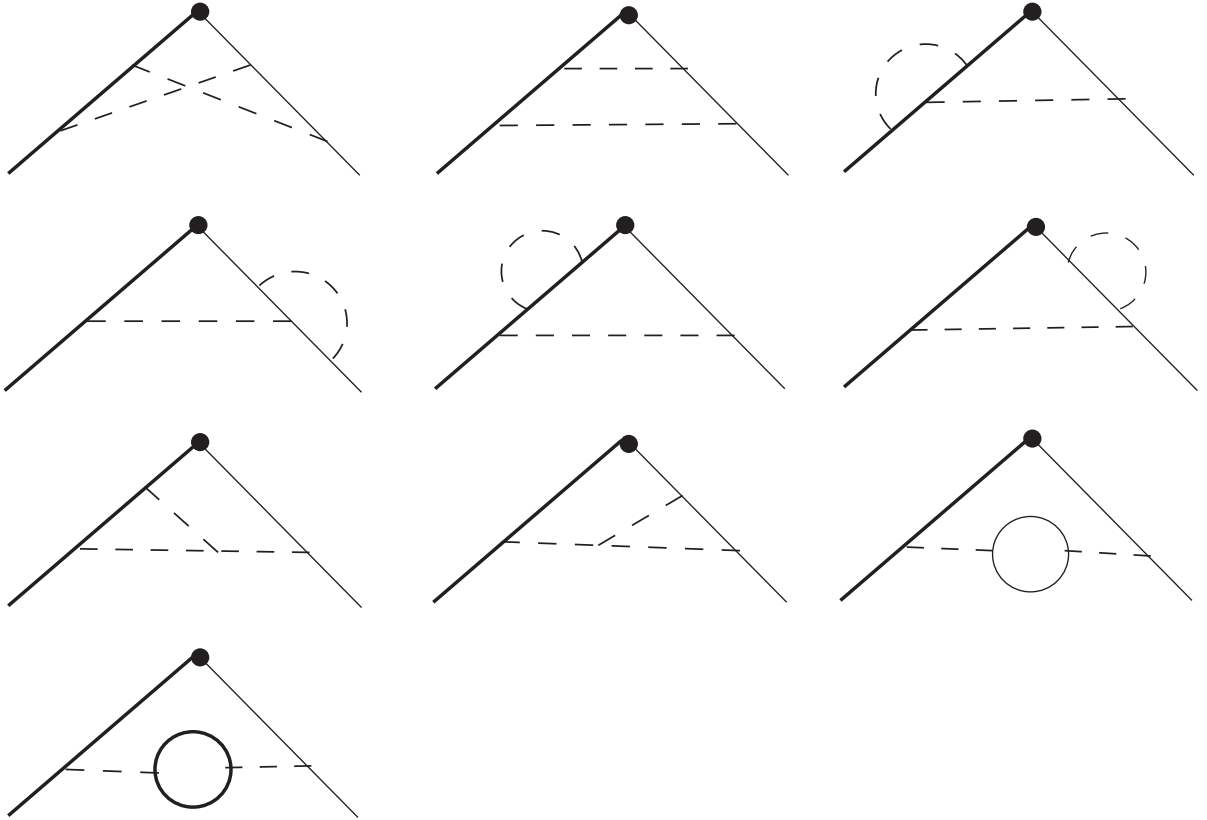, width=16cm}\hspace{3mm}%
\caption{\label{fig:diagrams}
Two-loop corrections to the $b \to u$ left-handed current.
The incoming $b$-quark and the outgoing $u$-quark are represented
by thick and thin solid lines, respectively, while dashed lines 
represent gluons. Fermionic bubbles with $b$-quarks and lighter
quarks (the latter being treated as massless) are shown by thick and thin
circles. Diagrams where the light fermionic bubbles are replaced
by gluons and ghost-particles are not shown explicitly, but they are
taken into account.}
\end{center}
\end{figure}

In this section we describe some technical details involved in 
obtaining the two-loop QCD corrections to the $b\to u$ current. 
The main task is to evaluate the bare two-loop amplitude
by calculating the Feynman diagrams in Figure 
\ref{fig:diagrams}. This bare amplitude contains both UV and IR divergences.
The UV divergences are removed by counterterms related to $b$ and $u$-quark
wave-function renormalization (on-shell scheme),
coupling constant renormalization ($\overline{\rm{MS}}$ scheme), 
and mass renormalization (on-shell scheme).

The calculation of the individual two-loop Feynman diagrams 
proceeds as follows.
First, by doing tensor decomposition, we 
extract the contributions of each diagram to the
form factors $D_i$ in (\ref{eq:DiDef}). At this
level, these contributions are written as  linear combinations of
certain scalar integrals.  Second, this
rather large set of scalar integrals is reduced to a much smaller set 
of master integrals using the Laporta algorithm  \cite{Laporta:2001dd},
which is based on the  integration-by-parts identities introduced
in   \cite{Tkachov:1981wb, Chetyrkin:1981qh}.  A very useful tool for
performing this reduction is the integral reduction program AIR
\cite{Anastasiou:2004vj}, written in \texttt{Maple}, and we have used this 
program in our calculation.   

A typical master integral depends on $m_b$, the dimensionless variable 
$\hat{s}=(p_b-p)^2/m_b^2$, and the 
parameter $\epsilon =(4-d)/2$ of dimensional regularization.  
Some of the simpler master integrals (those with three or less propagators),
are easily solved using the standard technique of Feynman parameterization.
In most cases, it is straightforward to obtain exact results in $\epsilon$,
which involve hypergeometric functions or their generalizations.  These can 
be expanded around $\epsilon \to 0$ using the \texttt{Mathematica} 
program \texttt{HypExp}
\cite{Huber:2005yg, Huber:2007dx}. 
For the more difficult master integrals, we have used the 
differential equation technique \cite{Remiddi:1997ny} 
(for a recent review, see \cite{Argeri:2007up}).
This involves solving a set of differential equations 
obtained by differentiating the master integrals with respect to the variable 
$\hat{s}$.  The solutions to the differential equations determine
the master integrals as a Laurent series in $\epsilon$, 
up to their values at the boundary point $\hat{s}=0$.  
In some cases, these constants can be determined by requiring that 
the coefficients in the Laurent expansion are finite 
in the limit $\hat{s} \to 0$.  In other cases, there is no choice 
but to calculate the $\epsilon$-expansion of the two-loop master 
integral at the point $\hat{s}=0$. The solutions to the 
differential equations involve 
the harmonic polylogarithms (HPLs) introduced in 
\cite{Remiddi:1999ew}.  For their numerical implementation and 
also some symbolic manipulations, we used
the \texttt{Mathematica} package \texttt{HPL} \cite{Maitre:2005uu}.

We have checked our results in several ways. First, we have used the 
numerical method of sector decomposition  \cite{Binoth:2003ak} to
evaluate the master integrals for various values of $\hat{s}$, and checked 
that they agree with the analytic results.  For this we have used 
self-written code, and also the publicly
available C++ program described in \cite{Bogner:2007cr}.
Second, we  have obtained 
results as a double series in $\epsilon \to 0$, $\hat{s} \to 0$ using two 
different techniques.  One is to expand each master integral as a series
in $\hat{s} \to 0$ before doing the loop integrals using sector decomposition, 
the other is to obtain results for each diagram at $\hat{s}=0$ and then recover
the $\hat{s}$-dependence using differential equations. We then checked that 
these agree numerically with the expansion of the analytic results in the 
same limit, up to the first five or six terms around $\hat{s} \to 0$.     
Finally, we were able to transform our basis of master integrals 
into that used for the two-loop calculation of the vertex corrections 
in $B\to \pi \pi$, presented in \cite{Bell:2006tz, Bell:2007tv}.
For some of the master integrals, we used these results to help 
convert numerical results for the boundary
conditions into analytic results in terms of constants like $\pi$.

To illustrate the method of differential equations in our application,
we take as an example the first diagram in the second row in Figure
\ref{fig:diagrams}. In this case we have four master integrals
$h_1$, $h_2$ , $h_3$ and $h_4$, reading
\begin{eqnarray}
h_1(\hat{s}) &=& \int \,  
\frac{d^d\ell}{(2\pi)^d} \,
\frac{d^dr}{(2\pi)^d} \, 
\frac{1}{[(\ell+p_b)^2-m_b^2]  \, [(l+r+p)^2] \, [(r+p)^2] } \, , \nonumber \\
h_2(\hat{s}) &=& \int \,  
\frac{d^d\ell}{(2\pi)^d} \,
\frac{d^dr}{(2\pi)^d} \, 
\frac{1}{[(\ell+p_b)^2-m_b^2]  \, [(\ell+p)^2] \,[(l+r+p)^2] \, [(r+p)^2] } \, , 
\nonumber \\
h_3(\hat{s}) &=& \int \,  
\frac{d^d\ell}{(2\pi)^d} \,
\frac{d^dr}{(2\pi)^d} \, 
\frac{1}{[(\ell+p_b)^2-m_b^2] \, [r^2] \, [(l+r+p)^2] } \, , \nonumber \\
h_4(\hat{s}) &=& \int \,  
\frac{d^d\ell}{(2\pi)^d} \,
\frac{d^dr}{(2\pi)^d} \, 
\frac{1}{[(\ell+p_b)^2-m_b^2] \, [r^2] \, [(\ell+p)^2]\, [(l+r+p)^2] } \, . 
\end{eqnarray}
They satisfy the differential equations 
\begin{eqnarray}
\frac{dh_1(\hat{s})}{d\hat{s}} &=& 0 \, ,
\nonumber \\
\frac{dh_2(\hat{s})}{d\hat{s}} &=& 
-\frac{1}{4} \, \frac{(4d + 4d \hat{s}-16\hat{s}-12)}{\hat{s}(1-\hat{s})} 
\, h_2(\hat{s}) -\frac{1}{4} \, \frac{(-3d+8)}{m_b^2 \, \hat{s}(1-\hat{s})} \, 
h_1(\hat{s}) \, ,
\nonumber \\
\frac{dh_3(\hat{s})}{d\hat{s}} &=& 
\frac{1}{4} \, \frac{(d-4)}{\hat{s}} 
\, h_3(\hat{s}) -\frac{1}{4} \, \frac{m_b^2 \, (1-\hat{s})(d-4)}{\hat{s}} \, 
h_4(\hat{s}) \, ,
\nonumber \\
\frac{dh_4(\hat{s})}{d\hat{s}} &=& 
-\frac{1}{4} \, \frac{(3d+5d\hat{s}-8-20\hat{s})}{\hat{s}(1-\hat{s})} 
\, h_4(\hat{s}) -\frac{1}{4} \, \frac{(-3d+8)}{m_b^2 \, \hat{s}(1-\hat{s})} \, 
h_3(\hat{s}) \, .
\end{eqnarray}
Obviously, $h_1$ has to be calculated using the standard technique
of Feynman parameterization. The dependence  of $h_2$ on $\hat{s}$
can then be determined by solving the second differential equation, 
in which $h_1$ plays the role of a given inhomogeneity. The requirement
that $h_2$ is non-singular for $\hat{s} \to 0$ uniquely determines the function
$h_2(\hat{s})$. The $\hat{s}$ dependence of the functions $h_3$ and $h_4$
can be obtained by solving the corresponding two differential equations
simultaneously (as an expansion in $\epsilon$). Specifying $h_3(\hat{s}=0)$
by means of standard Feynman parameterization and imposing the additional 
requirement
that $h_4(\hat{s})$ is non-singular for $\hat{s} \to 0$, uniquely determines
$h_3(\hat{s})$ and $h_4(\hat{s})$. 

\subsection{Renormalized scalar form factors}
\label{subsec:QCDresults}
We now give results for the UV-renormalized form factors in (\ref{eq:DiDef}),
which we expand in $\alpha_s$ according to 
\[
D_i = \delta_{i1} + \frac{\alpha_s}{4 \pi} \, D_i^{(1)} +
                    \left(\frac{\alpha_s}{4 \pi}\right)^2 \, D_i^{(2)} +
\ldots \, .
\]  
We start by listing the results of the one-loop contributions.  
To this end we further decompose the quantities $D_i^{(1)}$ as
\begin{equation}
\label{eq:oneloopsplitting}
D_i^{(1)}=C_F\left[\frac{R_{(-2),i}^{(1)}}{\epsilon^2}+
\frac{R_{(-1),i}^{(1)}}{\epsilon}+R_{(0),i}^{(1)}+
R_{(1),i}^{(1)} \epsilon+R_{(2),i}^{(1)} \epsilon^2 \right] \, .
\end{equation}
The Laurent expansion coefficients 
of the poles and constant term have been known for 
some time \cite{Bauer:2000yr}, 
whereas the terms proportional to $\epsilon$ and $\epsilon^2$ are new.
Note that terms up to $\epsilon^2$ are needed to correctly 
extract the Wilson coefficients $C_i$ through 
(\ref{eq:Matching}).  
The explicit results for the $R^{(i)}$ in 
(\ref{eq:oneloopsplitting}) read (recall $\hat s =(p_b-p)^2/m_b^2$)
\begin{eqnarray}
\nonumber
R_{(-2),1}^{(1)}&=&-1 \\ \nonumber
R_{(-1),1}^{(1)}&=&-\frac{5}{2}-2 L-2 F_4
\\\nonumber
R_{(0),1}^{(1)}&=&-6-5 L-2 L^2-\frac{\pi ^2}{12}-3
   F_4-4 L F_4+\frac{F_4}{\hat{s}}-2
   F_5-4 F_{10}
\\\nonumber
R_{(1),1}^{(1)}&=&-12-12 L-5 L^2-\frac{4
   L^3}{3}-\frac{5 \pi
   ^2}{24}-\frac{L \pi ^2}{6}-8 F_4-6
   L F_4-4 L^2 F_4-\frac{\pi ^2
   F_4}{6}+
\\&&\nonumber      
   \frac{4
   F_4}{\hat{s}}+
   \frac{2 L
   F_4}{\hat{s}}-3 F_5-4 L
   F_5+\frac{F_5}{\hat{s}}-2 F_6-6
   F_{10}-8 L F_{10}+\frac{2
   F_{10}}{\hat{s}}-4 F_{11}-
\\&&\nonumber         
   4 F_{13}-8
   F_{17}+\frac{\zeta(3)}{3}
   \\\nonumber
R_{(2),1}^{(1)}&=&
-24-24 L-12 L^2-\frac{10
   L^3}{3}-\frac{2 L^4}{3}-\frac{\pi
   ^2}{2}-\frac{5 L \pi
   ^2}{12}-\frac{L^2 \pi
   ^2}{6}-\frac{\pi ^4}{160}-16   F_4-
\\&&\nonumber            
   16 L F_4-6 L^2 F_4-\frac{8 L^3
   F_4}{3}-\frac{\pi ^2
   F_4}{4}-\frac{1}{3} L \pi ^2
   F_4+\frac{8 F_4}{\hat{s}}+\frac{8
   L F_4}{\hat{s}}+\frac{2 L^2
   F_4}{\hat{s}}+
\\&&\nonumber            
   \frac{\pi ^2 F_4}{12
   \hat{s}}-8 F_5-6 L F_5-4 L^2
   F_5-\frac{\pi ^2 F_5}{6}+\frac{4
   F_5}{\hat{s}}+\frac{2 L
   F_5}{\hat{s}}-3 F_6-4 L
   F_6+
   \\&&\nonumber         
   \frac{F_6}{\hat{s}}-2 F_7-16
   F_{10}-12 L F_{10}-8 L^2
   F_{10}-\frac{\pi ^2
   F_{10}}{3}+\frac{8
   F_{10}}{\hat{s}}+\frac{4 L
   F_{10}}{\hat{s}}-6 F_{11}-
   \\&&\nonumber         
   8 L
   F_{11}+\frac{2 F_{11}}{\hat{s}}-4
   F_{12}-6 F_{13}-8 L F_{13}+\frac{2
   F_{13}}{\hat{s}}-4 F_{14}-4
   F_{15}-12 F_{17}-
   \\&&\nonumber         
   16 L
   F_{17}+\frac{4 F_{17}}{\hat{s}}-8
   F_{18}-8 F_{19}-8 F_{20}-16
   F_{21}+\frac{5
   \zeta(3)}{6}+\frac{2}{3} L
   \zeta(3)+\frac{2}{3} F_4
   \zeta(3)
\end{eqnarray}
\begin{eqnarray}
\nonumber
R_{(-2),2}^{(1)}&=&R_{(-1),2}^{(1)}=0 \\ \nonumber
R_{(0),2}^{(1)}&=& \frac{2}{\hat{s}}-\frac{2F_4}{\hat{s}^2}+\frac{2F_4}{\hat{s}}
\\\nonumber         
R_{(1),2}^{(1)}&=&\frac{4}{\hat{s}}+\frac{4
   L}{\hat{s}}-\frac{2
   F_4}{\hat{s}^2}-\frac{4 L
   F_4}{\hat{s}^2}+\frac{2
   F_4}{\hat{s}}+\frac{4 L
   F_4}{\hat{s}}-\frac{2
   F_5}{\hat{s}^2}+\frac{2
   F_5}{\hat{s}}-\frac{4
   F_{10}}{\hat{s}^2}+\frac{4
   F_{10}}{\hat{s}}
   \\\nonumber         
R_{(2),2}^{(1)}&=&\frac{8}{\hat{s}}+\frac{8
   L}{\hat{s}}+\frac{4
   L^2}{\hat{s}}+\frac{\pi ^2}{6
   \hat{s}}-\frac{4
   F_4}{\hat{s}^2}-\frac{4 L
   F_4}{\hat{s}^2}-\frac{4 L^2
   F_4}{\hat{s}^2}-
   \frac{\pi ^2
   F_4}{6 \hat{s}^2}+
   \frac{4
   F_4}{\hat{s}}+\frac{4 L
   F_4}{\hat{s}}+
   \\&&\nonumber   
   \frac{4 L^2
   F_4}{\hat{s}}+
   \frac{\pi ^2 F_4}{6
   \hat{s}}-\frac{2
   F_5}{\hat{s}^2}-\frac{4 L
   F_5}{\hat{s}^2}+\frac{2
   F_5}{\hat{s}}+\frac{4 L
   F_5}{\hat{s}}-\frac{2
   F_6}{\hat{s}^2}+\frac{2
   F_6}{\hat{s}}-\frac{4
   F_{10}}{\hat{s}^2}-
   \\&&\nonumber   
   \frac{8 L
   F_{10}}{\hat{s}^2}+\frac{4
   F_{10}}{\hat{s}}+\frac{8 L
   F_{10}}{\hat{s}}-\frac{4
   F_{11}}{\hat{s}^2}+\frac{4
   F_{11}}{\hat{s}}-\frac{4
   F_{13}}{\hat{s}^2}+\frac{4
   F_{13}}{\hat{s}}-\frac{8
   F_{17}}{\hat{s}^2}+\frac{8
   F_{17}}{\hat{s}}
\end{eqnarray}
\begin{eqnarray}
\nonumber
R_{(-2),3}^{(1)}&=&R_{(-1),3}^{(1)}=0 \\ \nonumber
R_{(0),3}^{(1)}&=&-\frac{2}{\hat{s}}+\frac{2
   F_4}{\hat{s}^2}-\frac{4
   F_4}{\hat{s}}
    \\\nonumber         
R_{(1),3}^{(1)}&=&-\frac{4}{\hat{s}}-\frac{4
   L}{\hat{s}}+\frac{2
   F_4}{\hat{s}^2}+\frac{4 L
   F_4}{\hat{s}^2}-\frac{10
   F_4}{\hat{s}}-\frac{8 L
   F_4}{\hat{s}}+\frac{2
   F_5}{\hat{s}^2}-\frac{4
   F_5}{\hat{s}}+\frac{4
   F_{10}}{\hat{s}^2}-\frac{8
   F_{10}}{\hat{s}}
    \\\nonumber         
R_{(2),3}^{(1)}&=&-\frac{8}{\hat{s}}-\frac{8
   L}{\hat{s}}-\frac{4
   L^2}{\hat{s}}-\frac{\pi ^2}{6
   \hat{s}}+\frac{4
   F_4}{\hat{s}^2}+\frac{4 L
   F_4}{\hat{s}^2}+\frac{4 L^2
   F_4}{\hat{s}^2}+\frac{\pi ^2
   F_4}{6 \hat{s}^2}-\frac{20
   F_4}{\hat{s}}-\frac{20 L
   F_4}{\hat{s}}-
   \\&&\nonumber   
   \frac{8 L^2
   F_4}{\hat{s}}-\frac{\pi ^2 F_4}{3
   \hat{s}}+\frac{2
   F_5}{\hat{s}^2}+\frac{4 L
   F_5}{\hat{s}^2}-\frac{10
   F_5}{\hat{s}}-\frac{8 L
   F_5}{\hat{s}}+\frac{2
   F_6}{\hat{s}^2}-\frac{4
   F_6}{\hat{s}}+\frac{4
   F_{10}}{\hat{s}^2}+
   \\&&\nonumber 
   \frac{8 L
   F_{10}}{\hat{s}^2}-
    \frac{20
   F_{10}}{\hat{s}}-
       \frac{16 L
   F_{10}}{\hat{s}}+\frac{4
   F_{11}}{\hat{s}^2}-\frac{8
   F_{11}}{\hat{s}}+\frac{4
   F_{13}}{\hat{s}^2}-\frac{8
   F_{13}}{\hat{s}}+\frac{8
   F_{17}}{\hat{s}^2}-\frac{16
   F_{17}}{\hat{s}}    
\end{eqnarray}
In these equations  $L=\ln \mu/m_b$, while
the quantities $F_{1},\ldots,F_{21}$ denote the following harmonic
polylogarithms:       
\begin{eqnarray}
\label{eq:HPLset}
F= && \left[ \text{HPL}(\{-2\},\hat{s}),\text{HPL}(\{-1\},1-\hat{s}),
\text{HPL}(\{-1\},\hat{s}),\text{HPL}(
   \{1\},\hat{s}),\text{HPL}(\{2\},\hat{s}),
   \nonumber
\right. \\&&\nonumber \left.     
  \text{HPL}(\{3\},\hat{s}), \text{HPL}(\{4\},\hat{s}
   ),\text{HPL}(\{-2,2\},\hat{s}),
   \text{HPL}(\{-1,2\},\hat{s}),\text{HPL}(\{1,1\},\hat{s}),
   \right. \\&&\nonumber \left.  
   \text{HPL}(\{1,2\},\hat{s}),\text{HPL}(\{1,3\},\hat{s}),\text{HPL}(
   \{2,1\},\hat{s}),\text{HPL}(\{2
   ,2\},\hat{s}),\text{HPL}(\{3,1\},\hat{s}),
   \right. \\&&\nonumber \left.  
   \text{HPL}(\{-1,0,0\},1-\hat{s}),\text{HPL}(\{1,1,1
   \},\hat{s}),\text{HPL}(\{1,1,2\},\hat{s}),\text{HPL}(\{1,2,1\}
   ,\hat{s}),
   \right. \\&& \left.  
   \text{HPL}(\{2,1,1\},
   \hat{s}),\text{HPL}(\{1,1,1,1\}
   ,\hat{s})\right] \, .
   \end{eqnarray}
We now turn to the order $\alpha_s^2$ contributions $D_i^{(2)}$, which we
decompose according to
\begin{equation}
D_i^{(2)}=C_F\left[\frac{R_{(-4),i}^{(2)}}{\epsilon^4}+
\frac{R_{(-3),i}^{(2)}}{\epsilon^3}+\frac{R_{(-2),i}^{(2)}}{\epsilon^2}+
\frac{R_{(-1),i}^{(2)}}{\epsilon}+R_{(0),i}^{(2)}\right] \, .
\end{equation}
The (infrared) singular pieces yield relatively compact expressions. We
find
\begin{eqnarray}
\nonumber R_{(-4),1}^{(2)}&=&
\frac{C_F}{2}\\
\nonumber R_{(-3),1}^{(2)}&=&
C_F\left(\frac{5}{2}+2 L+2F_4\right)+\frac{11 C_A}{4}-n_l T_R\\
   \nonumber R_{(-2),1}^{(2)}&=&
C_F
   \left(\frac{73}{8}+10 L+4
   L^2+\frac{\pi ^2}{12}+8 F_4+8 L
   F_4-\frac{F_4}{\hat{s}}+2 F_5+8
   F_{10}\right)+
   \\&&\nonumber
   C_A \left(\frac{49}{18}+\frac{11
   L}{3}+\frac{\pi ^2}{12}+\frac{11
   F_4}{3}\right)+\frac{8}{3} L n_h
   T_R+\left(-\frac{10}{9}-\frac{4
   L}{3}-\frac{4 F_4}{3}\right) n_l
   T_R\\
   \nonumber R_{(-1),1}^{(2)}&=&
  C_F
   \left(\frac{213}{8}-\frac{19
   \text{$\zeta $(3)}}{3}+\frac{73
   L}{2}+20 L^2+\frac{16
   L^3}{3}+\frac{11 \pi
   ^2}{12}+\frac{L \pi
   ^2}{3}+\frac{55 F_4}{2}+32 L
   F_4+
   \right.\\&&\nonumber\left.
   16 L^2 F_4+\frac{\pi ^2
   F_4}{3}-\frac{13 F_4}{2
   \hat{s}}-\frac{4 L F_4}{\hat{s}}+8
   F_5+8 L F_5-\frac{F_5}{\hat{s}}+2
   F_6+28 F_{10}+
\right.\\&&\nonumber\left.   
   32 L F_{10}-\frac{6
   F_{10}}{\hat{s}}+8 F_{11}+12
   F_{13}+32
   F_{17}\right)+C_A \left(-\frac{1549}{216}+\frac{11
   \text{$\zeta $(3)}}{2}-\frac{67
   L}{9}-
   \right.\\&&\nonumber\left.
   \frac{7 \pi ^2}{24}+\frac{L
   \pi ^2}{3}-\frac{67
   F_4}{9}+\frac{\pi ^2
   F_4}{3}\right)+\left(\frac{20
   L}{3}+8 L^2+\frac{\pi
   ^2}{9}+\frac{16 L F_4}{3}\right)
   n_h
   T_R+
   \\&&\nonumber
   \left(\frac{125}{54}+\frac{20
   L}{9}+\frac{\pi ^2}{6}+\frac{20
   F_4}{9}\right) n_l T_R 
\end{eqnarray}
\begin{eqnarray}
   R_{(-4),2}^{(2)}&=&R_{(-3),2}^{(2)}=0\nonumber
\\\nonumber 
R_{(-2),2}^{(2)}&=&C_F \left(-\frac{2}{\hat{s}}+\frac{2
   F_4}{\hat{s}^2}-\frac{2
   F_4}{\hat{s}}\right)
\\\nonumber 
R_{(-1),2}^{(2)}&=&C_F \left(-\frac{9}{\hat{s}}-\frac{8
   L}{\hat{s}}+\frac{7
   F_4}{\hat{s}^2}+\frac{8 L
   F_4}{\hat{s}^2}-\frac{11
   F_4}{\hat{s}}-\frac{8 L
   F_4}{\hat{s}}+\frac{2
   F_5}{\hat{s}^2}-\frac{2
   F_5}{\hat{s}}+\frac{12
   F_{10}}{\hat{s}^2}-\frac{12
   F_{10}}{\hat{s}}\right)
\end{eqnarray}
\begin{eqnarray}
R_{(-4),3}^{(2)}&=&R_{(-3),3}^{(2)}=0\nonumber
\\\nonumber 
R_{(-2),3}^{(2)}&=&C_F \left(\frac{2}{\hat{s}}-\frac{2
   F_4}{\hat{s}^2}+\frac{4
   F_4}{\hat{s}}\right)
\\\nonumber
R_{(-1),3}^{(2)}&=&C_F \left(\frac{9}{\hat{s}}+\frac{8
   L}{\hat{s}}-\frac{7
   F_4}{\hat{s}^2}-\frac{8 L
   F_4}{\hat{s}^2}+\frac{24
   F_4}{\hat{s}}+\frac{16 L
   F_4}{\hat{s}}-\frac{2
   F_5}{\hat{s}^2}+\frac{4
   F_5}{\hat{s}}-\frac{12
   F_{10}}{\hat{s}^2}+\frac{24
   F_{10}}{\hat{s}}\right) \, .
   \end{eqnarray}   
On the other hand, the expressions for the 
infrared finite parts $R^{(2)}_{(0),i}$ 
are rather lengthy. It is convenient to further decompose them
according to
\begin{eqnarray}
R_{(0),i}^{(2)}=\sum_{j,k}^{} \frac{C_Ff_{i,j,k}^{\text{a}}+C_Af_{i,j,k}^{\text{na}}+
n_lT_Rf_{i,j,k}^{\text{nl}}+n_hT_Rf_{i,j,k}^{\text{nh}}}
{\hat{s}^j(1-\hat{s})^k} \, .
\nonumber
\end{eqnarray}
In the following we list the
functions $f_{i,j,k}^{\text{a}}$, $f_{i,j,k}^{\text{na}}$,
$f_{i,j,k}^{\text{nl}}$ and 
$f_{i,j,k}^{\text{nh}}$, $(i=1,2,3)$ 
for all values $j,k$ for which they are nonzero. We find
\begin{eqnarray}
\nonumber
f_{1,0,0}^{a}&=&\frac{1327}{16}+\frac{16 \text{$\zeta
   $(3)}}{3}+\frac{213 L}{2}-\frac{76
   \text{$\zeta $(3)} L}{3}+73
   L^2+\frac{80 L^3}{3}+\frac{16
   L^4}{3}+\frac{97 \pi ^2}{48}-4
   \text{ln(2)} \pi ^2+
   \\&&\nonumber
   \frac{11 L \pi
   ^2}{3}+\frac{2 L^2 \pi
   ^2}{3}-\frac{449 \pi
   ^4}{720}-\frac{4 \pi ^2
   F_1}{3}+\frac{10 \pi ^2
   F_3}{3}+\frac{153 F_4}{2}-\frac{28
   \text{$\zeta $(3)} F_4}{3}+110 L
   F_4+
\\&&\nonumber   
   64 L^2 F_4+\frac{64 L^3
   F_4}{3}+\frac{10 \pi ^2
   F_4}{3}+\frac{4}{3} L \pi ^2
   F_4-\frac{19 F_5}{2}+32 L F_5+16
   L^2 F_5+\frac{\pi ^2 F_5}{3}-
\\&&\nonumber   
   12 F_6+8 L F_6-6 F_7-16 F_8+40 F_9+59
   F_{10}+112 L F_{10}+64 L^2
   F_{10}+\frac{4 \pi ^2
   F_{10}}{3}+
\\&&\nonumber      
   28 F_{11}+32 L F_{11}-8
   F_{12}+60 F_{13}+48 L F_{13}+12
   F_{14}+12 F_{15}+104 F_{17}+128 L
   F_{17}+
\\&&\nonumber         
   32 F_{18}+48 F_{19}+56
   F_{20}+128 F_{21}
\\\nonumber
f_{1,1,0}^{a}&=&\frac{2 \pi ^2 F_3}{3}-\frac{49
   F_4}{2}-26 L F_4-8 L^2 F_4+\frac{5
   \pi ^2 F_4}{6}-\frac{15 F_5}{2}-4
   L F_5+F_6+8 F_9-
\\&&\nonumber            
   25 F_{10}-24 L
   F_{10}-4 F_{11}-10 F_{13}-28
   F_{17}
   \\\nonumber
f_{1,0,1}^{a}&=&-30 \text{$\zeta $(3)}+\frac{28 \pi
   ^2}{3}+16 \text{ln(2)} \pi
   ^2+\frac{3 \pi ^4}{5}+\pi ^2
   F_2-\frac{20 \pi ^2
   F_3}{3}+\frac{28 \pi ^2 F_4}{3}+90
   F_5-
\\&&\nonumber               
   4 \pi ^2 F_5+12 F_6+8 F_7-80
   F_9+50 F_{10}+24 F_{11}-78
   F_{13}-8 F_{14}+16 F_{15}+2 F_{16}
\\\nonumber
f_{1,0,2}^{a}&=&-\frac{59 \pi ^2}{3}-\frac{277 \pi
   ^4}{90}-\frac{8 \pi ^2
   F_1}{3}+\frac{8 \pi ^2
   F_3}{3}-\frac{68 \pi ^2 F_4}{3}-50
   F_5+\frac{62 \pi ^2 F_5}{3}+24
   F_6-
\\&&\nonumber                  
   20 F_7-32 F_8+32 F_9-56
   F_{11}+112 F_{13}+52 F_{14}-104
   F_{15}
\\\nonumber
f_{1,0,3}^{a}&=&\frac{152 \pi ^4}{45}+\frac{8 \pi ^2
   F_1}{3}-3 \pi ^2 F_2-\frac{68 \pi
   ^2 F_5}{3}+6 F_6+24 F_7+32 F_8-6
   F_{13}-56 F_{14}+
\\&&\nonumber                     
   112 F_{15}-6
   F_{16}
\end{eqnarray}
\begin{eqnarray}
\nonumber
f_{1,0,0}^{na}&=&-\frac{89437}{1296}+\frac{19
   \text{$\zeta $(3)}}{18}-\frac{3925
   L}{54}+22 \text{$\zeta $(3)}
   L-\frac{299 L^2}{9}-\frac{44
   L^3}{9}-\frac{815 \pi ^2}{216}+2
   \text{ln(2)} \pi ^2-
\\&&\nonumber                        
   \frac{16 L \pi
   ^2}{9}+\frac{2 L^2 \pi
   ^2}{3}+\frac{31 \pi
   ^4}{120}+\frac{2 \pi ^2
   F_1}{3}-\frac{5 \pi ^2
   F_3}{3}-\frac{2545 F_4}{54}+14
   \text{$\zeta $(3)} F_4-\frac{466 L
   F_4}{9}-
\\&&\nonumber                           
   \frac{44 L^2
   F_4}{3}-\frac{28 \pi ^2
   F_4}{9}+\frac{4}{3} L \pi ^2
   F_4-\frac{116 F_5}{9}-\frac{44 L
   F_5}{3}+\frac{4 \pi ^2
   F_5}{3}+\frac{20 F_6}{3}+8 F_8-
   \\&&\nonumber                        
   20F_9-\frac{349 F_{10}}{9}-\frac{88
   L F_{10}}{3}+\frac{4 \pi ^2
   F_{10}}{3}-\frac{44 F_{11}}{3}+8
   F_{12}-\frac{62
   F_{13}}{3}-\frac{88 F_{17}}{3}
   \\\nonumber
f_{1,1,0}^{na}&=&-\frac{1}{3} \pi ^2 F_3+\frac{269
   F_4}{18}+\frac{22 L
   F_4}{3}-\frac{2 \pi ^2
   F_4}{3}+\frac{11 F_5}{3}-4
   F_9+\frac{13 F_{10}}{3}
   \\\nonumber
f_{1,0,1}^{na}&=&15 \text{$\zeta $(3)}+13 \pi ^2-8
   \text{ln(2)} \pi ^2+\frac{3 \pi
   ^4}{5}-\frac{\pi ^2
   F_2}{2}+\frac{10 \pi ^2
   F_3}{3}+\frac{47 \pi ^2 F_4}{6}+12
   F_5-
\\&&\nonumber   
   4 \pi ^2 F_5-31 F_6+8 F_7+40
   F_9+17 F_{10}+13 F_{11}-11
   F_{13}-8 F_{14}+16 F_{15}-F_{16}
   \\\nonumber
f_{1,0,2}^{na}&=&-\frac{67 \pi ^2}{6}-\frac{86 \pi
   ^4}{45}+\frac{4 \pi ^2
   F_1}{3}-\frac{4 \pi ^2
   F_3}{3}-\frac{29 \pi ^2 F_4}{3}-17
   F_5+\frac{38 \pi ^2 F_5}{3}+30
   F_6-
   \\&&\nonumber 
 36 F_7+16 F_8-16 F_9-14
   F_{11}+28 F_{13}+20 F_{14}-40
   F_{15}
      \\\nonumber
      f_{1,0,3}^{na}&=&  \frac{263 \pi ^4}{180}-\frac{4 \pi ^2
   F_1}{3}+\frac{3 \pi ^2
   F_2}{2}-\frac{29 \pi ^2 F_5}{3}-3
   F_6+30 F_7-16 F_8+3 F_{13}-14
   F_{14}+
\\&&\nonumber    
   28 F_{15}+3 F_{16}
\end{eqnarray}
\begin{eqnarray} \nonumber
      f_{1,0,0}^{nl}&=&\frac{6629}{324}+\frac{26
   \text{$\zeta $(3)}}{9}+\frac{682
   L}{27}+\frac{100 L^2}{9}+\frac{16
   L^3}{9}+\frac{85 \pi
   ^2}{54}+\frac{8 L \pi
   ^2}{9}+\frac{418
   F_4}{27}+
\\&&\nonumber    
   \frac{152 L
   F_4}{9}+\frac{16 L^2
   F_4}{3}+\frac{8 \pi ^2
   F_4}{9}+\frac{76 F_5}{9}+\frac{16
   L F_5}{3}+\frac{8
   F_6}{3}+\frac{152
   F_{10}}{9}+\frac{32 L
   F_{10}}{3}+
\\&&\nonumber
   \frac{16F_{11}}{3}+\frac{16
   F_{13}}{3}+\frac{32 F_{17}}{3}
 \\\nonumber
      f_{1,1,0}^{nl}&=&-\frac{38 F_4}{9}-\frac{8 L
   F_4}{3}-\frac{4 F_5}{3}-\frac{8
   F_{10}}{3}
\end{eqnarray}
\begin{eqnarray} \nonumber
      f_{1,0,0}^{nh}&=&\frac{7951}{162}-\frac{28
   \text{$\zeta $(3)}}{9}+16 L+20
   L^2+\frac{112 L^3}{9}-\frac{41 \pi
   ^2}{54}+\frac{2 L \pi
   ^2}{3}+\frac{530 F_4}{27}+8 L
   F_4+
\\&&\nonumber   
   16 L^2 F_4+\frac{2 \pi ^2
   F_4}{9}-\frac{76 F_5}{9}+\frac{16
   L F_5}{3}+\frac{8 F_6}{3}+\frac{32
   L F_{10}}{3}
   \\\nonumber
      f_{1,1,0}^{nh}&=&-\frac{38 F_4}{9}-\frac{8 L
   F_4}{3}-\frac{4 F_5}{3}
    \\\nonumber
      f_{1,0,1}^{nh}&=&-\frac{508}{9}-\frac{64 \pi
   ^2}{9}-\frac{440 F_4}{9}+\frac{104
   F_5}{3}
     \\\nonumber
      f_{1,0,2}^{nh}&=&\frac{128}{9}+16 \text{$\zeta
   $(3)}+\frac{32 \pi
   ^2}{3}+\frac{128 F_4}{9}-48 F_5-16
   F_6
     \\\nonumber
      f_{1,0,3}^{nh}&=&-16 \text{$\zeta $(3)}-\frac{64 \pi
   ^2}{27}+\frac{128 F_5}{9}+16 F_6
\end{eqnarray}
\begin{eqnarray} \nonumber
      f_{2,1,0}^{a}&=&-31-36 L-16 L^2+3 \pi ^2+\frac{4 \pi
   ^2 F_3}{3}-24 F_4-44 L F_4-16 L^2
   F_4-\frac{19 \pi ^2 F_4}{3}+
\\&&\nonumber      
   23 F_5-8 L F_5+18 F_6+16 F_9+50
   F_{10}-48 L F_{10}-24 F_{11}+12
   F_{13}-56 F_{17}
    \\\nonumber
      f_{2,2,0}^{a}&=&-\frac{4}{3} \pi ^2 F_3+16 F_4+28 L
   F_4+16 L^2 F_4-\frac{5 \pi ^2
   F_4}{3}+13 F_5+8 L F_5-2 F_6-16
   F_9-
\\&&\nonumber         
   2 F_{10}+48 L F_{10}+8
   F_{11}+20 F_{13}+56 F_{17}
   \\\nonumber
      f_{2,3,0}^{a}&=&8 F_{10}
       \\\nonumber
      f_{2,0,1}^{a}&=& -4 \text{$\zeta $(3)}+\frac{28 \pi
   ^2}{3}-2 \pi ^2 F_2+\frac{8 \pi ^2
   F_3}{3}-\frac{20 \pi ^2 F_4}{3}+44
   F_5+32 F_9+100 F_{10}-
\\&&\nonumber            
   24 F_{11}+44
   F_{13}-4 F_{16}
   \\\nonumber
      f_{2,0,2}^{a}&=&32 \text{$\zeta $(3)}-\frac{118 \pi
   ^2}{3}-\frac{12 \pi ^4}{5}-8 \pi
   ^2 F_2+\frac{16 \pi ^2
   F_3}{3}-\frac{136 \pi ^2
   F_4}{3}-100 F_5+16 \pi ^2 F_5+
\\&&\nonumber               
   32   F_6-32 F_7+64 F_9-112 F_{11}+208
   F_{13}+32 F_{14}-64 F_{15}-16
   F_{16}
   \\\nonumber  
   f_{2,0,3}^{a}&=&\frac{304 \pi ^4}{45}+\frac{16 \pi ^2
   F_1}{3}-6 \pi ^2 F_2-\frac{136 \pi
   ^2 F_5}{3}+12 F_6+48 F_7+64 F_8-12
   F_{13}-
\\&&\nonumber                  
   112 F_{14}+224 F_{15}-12
   F_{16}
\end{eqnarray}
\begin{eqnarray}
\nonumber
   f_{2,1,0}^{na}&=&\frac{269}{9}+\frac{44 L}{3}-2 \pi
   ^2-\frac{2 \pi ^2
   F_3}{3}+\frac{257 F_4}{9}+\frac{44
   L F_4}{3}-\frac{10 \pi ^2
   F_4}{3}+\frac{46 F_5}{3}+4 F_6-8
   F_9+
\\&&\nonumber                     
   \frac{86 F_{10}}{3}-4 F_{11}+8
   F_{13}
   \\\nonumber
   f_{2,2,0}^{na}&=&\frac{2 \pi ^2 F_3}{3}-\frac{203
   F_4}{9}-\frac{44 L F_4}{3}+\frac{4
   \pi ^2 F_4}{3}-\frac{22 F_5}{3}+8
   F_9-\frac{26 F_{10}}{3}
   \\\nonumber
   f_{2,0,1}^{na}&=&2 \text{$\zeta $(3)}+\frac{2 \pi
   ^2}{3}+\pi ^2 F_2-\frac{4 \pi ^2
   F_3}{3}-\frac{5 \pi ^2 F_4}{3}+32
   F_5+10 F_6-16 F_9+34 F_{10}+
\\&&\nonumber                        
   2 F_{11}-2 F_{13}+2 F_{16}
   \\\nonumber
   f_{2,0,2}^{na}&=&-16 \text{$\zeta $(3)}-\frac{67 \pi
   ^2}{3}-\frac{6 \pi ^4}{5}+4 \pi ^2
   F_2-\frac{8 \pi ^2
   F_3}{3}-\frac{58 \pi ^2 F_4}{3}-34
   F_5+8 \pi ^2 F_5+
\\&&\nonumber                           
   68 F_6-16 F_7-32
   F_9-28 F_{11}+64 F_{13}+16
   F_{14}-32 F_{15}+8 F_{16}
 \\\nonumber
   f_{2,0,3}^{na}&=&\frac{263 \pi ^4}{90}-\frac{8 \pi ^2
   F_1}{3}+3 \pi ^2 F_2-\frac{58 \pi
   ^2 F_5}{3}-6 F_6+60 F_7-32 F_8+6
   F_{13}-28 F_{14}+
\\&&\nonumber                              
   56 F_{15}+6
   F_{16}
\end{eqnarray}
\begin{eqnarray} \nonumber
   f_{2,1,0}^{nl}&=&-\frac{76}{9}-\frac{16 L}{3}-\frac{52
   F_4}{9}-\frac{16 L F_4}{3}-\frac{8
   F_5}{3}-\frac{16 F_{10}}{3}
     \\\nonumber
   f_{2,2,0}^{nl}&=&\frac{52 F_4}{9}+\frac{16 L
   F_4}{3}+\frac{8 F_5}{3}+\frac{16
   F_{10}}{3}
\end{eqnarray}
\begin{eqnarray} \nonumber
   f_{2,1,0}^{nh}&=&-\frac{76}{9}-\frac{16
   L}{3}-\frac{292 F_4}{9}-\frac{16 L
   F_4}{3}-\frac{8 F_5}{3}
    \\\nonumber
   f_{2,2,0}^{nh}&=&\frac{52 F_4}{9}+\frac{16 L
   F_4}{3}+\frac{8 F_5}{3}
     \\\nonumber
   f_{2,0,1}^{nh}&=&-\frac{104}{3}+\frac{32 \pi
   ^2}{9}-\frac{80 F_4}{3}-\frac{16
   F_5}{3}
     \\\nonumber
   f_{2,0,2}^{nh}&=&\frac{64 \pi ^2}{9}-\frac{32 F_5}{3}
     \\\nonumber
   f_{2,0,3}^{nh}&=&-32 \text{$\zeta $(3)}+32 F_6
\end{eqnarray}
\begin{eqnarray} \nonumber
   f_{3,1,0}^{a}&=&31+36 L+16 L^2-3 \pi ^2-\frac{8 \pi
   ^2 F_3}{3}+75 F_4+96 L F_4+32 L^2
   F_4+\frac{14 \pi ^2 F_4}{3}-
\\&&\nonumber                                 
   12 F_5+16 L F_5-20 F_6-32 F_9+96 L
   F_{10}+32 F_{11}+8 F_{13}+112
   F_{17}
    \\\nonumber
   f_{3,2,0}^{a}&=&\frac{4 \pi ^2 F_3}{3}-16 F_4-28 L
   F_4-16 L^2 F_4+\frac{5 \pi ^2
   F_4}{3}-13 F_5-8 L F_5+2 F_6+16
   F_9+
\\&&\nonumber                                    
   2 F_{10}-48 L F_{10}-8
   F_{11}-20 F_{13}-56 F_{17}
    \\\nonumber
   f_{3,3,0}^{a}&=&-8 F_{10}
    \\\nonumber
   f_{3,0,1}^{a}&=&-40 \text{$\zeta $(3)}-\frac{16 \pi
   ^2}{3}+16 \text{ln(2)} \pi ^2+4
   \pi ^2 F_2-\frac{16 \pi ^2
   F_3}{3}-4 F_4+\frac{28 \pi ^2
   F_4}{3}+12 F_5-
\\&&\nonumber                                       
   8 F_6-64 F_9-48
   F_{10}+8 F_{11}-56 F_{13}+8 F_{16}
    \\\nonumber
   f_{3,0,2}^{a}&=&12 \text{$\zeta $(3)}-\frac{248 \pi
   ^2}{3}-\frac{12 \pi ^4}{5}+6 \pi
   ^2 F_2+\frac{8 \pi ^2
   F_3}{3}-\frac{116 \pi ^2
   F_4}{3}-284 F_5+16 \pi ^2 F_5+
   \\&&\nonumber    
   32
   F_6-32 F_7+                                      
   32 F_9-300 F_{10}-88
   F_{11}+188 F_{13}+32 F_{14}-64
   F_{15}+12 F_{16}
    \\\nonumber
   f_{3,0,3}^{a}&=&-64 \text{$\zeta $(3)}+118 \pi
   ^2+\frac{716 \pi ^4}{45}+\frac{32
   \pi ^2 F_1}{3}+16 \pi ^2 F_2-16
   \pi ^2 F_3+136 \pi ^2 F_4+
    \\&&\nonumber    
   300 F_5-
   \frac{320 \pi ^2 F_5}{3}-112
   F_6+128 F_7+128 F_8-192 F_9+336
   F_{11}-
    \\&&\nonumber    
   640 F_{13}-256 F_{14}+512
   F_{15}+32 F_{16}
\\\nonumber
   f_{3,0,4}^{a}&=&-\frac{304 \pi ^4}{15}-16 \pi ^2
   F_1+18 \pi ^2 F_2+136 \pi ^2
   F_5-36 F_6-144 F_7-192 F_8+36
   F_{13}+
 \\&&\nonumber       
   336 F_{14}-672 F_{15}+36
   F_{16}
\end{eqnarray}
\begin{eqnarray} \nonumber
   f_{3,1,0}^{na}&=&-\frac{269}{9}-\frac{44 L}{3}+2 \pi
   ^2+\frac{4 \pi ^2
   F_3}{3}-\frac{592 F_4}{9}-\frac{88
   L F_4}{3}+\frac{14 \pi ^2
   F_4}{3}-\frac{68 F_5}{3}-4 F_6+
 \\&&\nonumber         
   16 F_9-\frac{112 F_{10}}{3}+4
   F_{11}-8 F_{13}
    \\\nonumber
   f_{3,2,0}^{na}&=&-\frac{2}{3} \pi ^2 F_3+\frac{203
   F_4}{9}+\frac{44 L F_4}{3}-\frac{4
   \pi ^2 F_4}{3}+\frac{22 F_5}{3}-8
   F_9+\frac{26 F_{10}}{3}
    \\\nonumber
   f_{3,0,1}^{na}&=&-6+20 \text{$\zeta $(3)}+\frac{38 \pi
   ^2}{3}-8 \text{ln(2)} \pi ^2-2 \pi
   ^2 F_2+\frac{8 \pi ^2 F_3}{3}-18
   F_4+\frac{16 \pi ^2 F_4}{3}-16
   F_6+
 \\&&\nonumber           
   32 F_9-4 F_{10}+16 F_{11}-12
   F_{13}-4 F_{16}
     \\\nonumber
   f_{3,0,2}^{na}&=&-6 \text{$\zeta $(3)}-\frac{118 \pi
   ^2}{3}-\frac{12 \pi ^4}{5}-3 \pi
   ^2 F_2-\frac{4 \pi ^2
   F_3}{3}-\frac{89 \pi ^2
   F_4}{3}-144 F_5+16 \pi ^2 F_5+
 \\&&\nonumber             
   82
   F_6-32 F_7-16 F_9-102 F_{10}-54
   F_{11}+102 F_{13}+32 F_{14}-64
   F_{15}-6 F_{16}
     \\\nonumber
   f_{3,0,3}^{na}&=&32 \text{$\zeta $(3)}+67 \pi
   ^2+\frac{398 \pi ^4}{45}-\frac{16
   \pi ^2 F_1}{3}-8 \pi ^2 F_2+8 \pi
   ^2 F_3+58 \pi ^2 F_4+102
   F_5-\frac{176 \pi ^2 F_5}{3}-
   \\&&\nonumber    
   196
   F_6+160 F_7-64 F_8+96 F_9+84
   F_{11}-184 F_{13}-96 F_{14}+192
   F_{15}-16 F_{16}
     \\\nonumber
   f_{3,0,4}^{na}&=&-\frac{263 \pi ^4}{30}+8 \pi ^2 F_1-9
   \pi ^2 F_2+58 \pi ^2 F_5+18
   F_6-180 F_7+96 F_8-18 F_{13}+84
   F_{14}-
   \\&&\nonumber 
   168 F_{15}-18 F_{16}
\end{eqnarray}
\begin{eqnarray} \nonumber
   f_{3,1,0}^{nl}&=&  \frac{76}{9}+\frac{16 L}{3}+\frac{152
   F_4}{9}+\frac{32 L
   F_4}{3}+\frac{16 F_5}{3}+\frac{32
   F_{10}}{3}
     \\\nonumber
   f_{3,2,0}^{nl}&=&-\frac{52 F_4}{9}-\frac{16 L
   F_4}{3}-\frac{8 F_5}{3}-\frac{16
   F_{10}}{3}
\end{eqnarray}
\begin{eqnarray} \nonumber
   f_{3,1,0}^{nh}&=&\frac{76}{9}+\frac{16 L}{3}+\frac{392
   F_4}{9}+\frac{32 L
   F_4}{3}+\frac{16 F_5}{3}
    \\\nonumber
   f_{3,2,0}^{nh}&=&-\frac{52 F_4}{9}-\frac{16 L
   F_4}{3}-\frac{8 F_5}{3}
    \\\nonumber
   f_{3,0,1}^{nh}&=&-\frac{32}{3}-\frac{16 \pi
   ^2}{9}+\frac{80 F_4}{3}+\frac{32
   F_5}{3}
   \\\nonumber
   f_{3,0,2}^{nh}&=&\frac{568}{3}+\frac{32 \pi
   ^2}{3}+\frac{496 F_4}{3}-48 F_5
\\\nonumber
   f_{3,0,3}^{nh}&=&-64 \text{$\zeta $(3)}-\frac{320 \pi
   ^2}{9}+\frac{352 F_5}{3}+64 F_6
\\\nonumber
   f_{3,0,4}^{nh}&=&96 \text{$\zeta $(3)}-96 F_6 \, .
\end{eqnarray}

\section{Two-loop results for the Wilson coefficients $C_i$}
\label{sec:Wilson}
In this section we give results for the Wilson 
coefficients $C_i$, valid to NNLO in $\alpha_s$. 
To calculate them, we take the 
UV-renormalized form factors 
$D_i$ obtained in the previous section, translate them to 
four-flavor QCD using (\ref{eq:decoupling}), and evaluate
the matching condition (\ref{eq:Matching}).  This procedure allows 
us to determine both the Wilson coefficients $C_i$ and the 
renormalization factor $Z_J$.
The form of the renormalization factor is completely determined by the 
renormalization-group equations for heavy-to-light currents in SCET, 
and thus provides important checks on our result.  We shall first 
say a few words about these, and then list results for the 
Wilson coefficients $C_i$.  

The renormalization factor $Z_J$ is determined from our calculation
by requiring that the matching relation (\ref{eq:Matching}) 
is finite in the limit $\epsilon \to 0$.  However, as explained 
in Section  \ref{sec:SCET}, it can also be determined by the UV poles
of the object $J\otimes S$ in the parton model.  Expressions at 
two loops can be  derived from the renormalization
factors for the jet and soft functions calculated in
 \cite{Becher:2006qw,Becher:2005pd}. Either way, the result 
depends only on logs
of the form $L_p\equiv \ln\mu/\bar n \cdot p$ and reads 
\begin{equation}
Z_J=1+\frac{C_F \alpha_s}{4\pi}\left[-\frac{1}{\epsilon^2}+
\frac{1}{\epsilon}\bigg(-\frac{5}{2}-2 L_p\bigg)\right]+
\left(\frac{\alpha_s}{4\pi}\right)^2 C_F 
\sum_{i=1}^4 \frac{Z_{(-i)}^{(2)}}{\epsilon^i}
\end{equation}
where the two-loop coefficients are
\begin{eqnarray}
\label{eq:zj}
Z_{(-4)}^{(2)}&=&\frac{C_F}{2} \nonumber \\
Z_{(-3)}^{(2)}&=& C_F\bigg(\frac{5}{2}+2 L_p\bigg)
+\frac{11 C_A}{4}- n_l T_R \nonumber \\
Z_{(-2)}^{(2)}&=&C_F \bigg(\frac{25}{8}+5 L_p+2 L_p^2\bigg)
+C_A\bigg(\frac{49}{18}+\frac{\pi^2}{12} +\frac{11 L_p}{3}\bigg)
+n_l T_R \bigg(-\frac{10}{9}-\frac{4 L_p}{3}\bigg) \nonumber \\
Z_{(-1)}^{(2)}&=& C_F\bigg(\frac{-3}{8}+\frac{\pi^2}{2}-6 \zeta_3   \bigg) 
+ C_A\bigg(-\frac{1549}{216}-\frac{7\pi^2}{24}+\frac{11}{2}\zeta_3
+L_p\bigg[-\frac{67}{9}+\frac{\pi^2}{3}\bigg]\bigg) \nonumber \\
&& +n_l T_R\bigg(\frac{125}{54}+\frac{\pi^2}{6}+\frac{20 L_p}{9}\bigg) \, .
\end{eqnarray}
In SCET, the hard function is derived from the matrix of 
Wilson coefficients $H_{ij}=C_i C_j$ and satisfies
the renormalization-group equation \cite{Bosch:2004th}
\begin{equation}
\label{eq:RGH}
\frac{d}{d\ln \mu} H_{ij}
(\bar n \cdot p, \mu)=
2\left[
\gamma^\prime(\alpha_s)+\Gamma_{\rm cusp}(\alpha_s)\ln\frac{\bar n \cdot p}{\mu}\right]H_{ij}(\bar n \cdot p,\mu) \,.
\end{equation} 
It is easy to show that the anomalous dimension derived from the 
explicit expressions in (\ref{eq:zj}) are
consistent with (\ref{eq:RGH}), with 
\begin{eqnarray}
\gamma'&=&-5\frac{ C_F\alpha_s}{4 \pi}
-8 C_F \left(\frac{\alpha_s}{4\pi}\right)^2
  \bigg[ C_F \left(\frac{3}{16} - \frac{\pi^2}{4}
    + 3\zeta_3 \right) 
    + C_A \left( \frac{1549}{432} + \frac{7\pi^2}{48} - \frac{11}{4}\,\zeta_3
    \right)\, \nonumber \\ &&
    - n_l T_R \left( \frac{125}{108} + \frac{\pi^2}{12} \right)  \bigg] \,.
\end{eqnarray}
This result is consistent with that given in \cite{Neubert:2004dd}, 
and the piece of the anomalous dimension 
proportional to the logarithmic term is consistent with the two-loop cusp 
anomalous dimension from \cite{Korchemskaya:1992je}.  

We now give final results for the Wilson coefficients $C_i$, which 
we decompose according to 
\begin{eqnarray}
C_i=C_i^{(0)}+\frac{\alpha_s}{4\pi} C_i^{(1)}+
\left(\frac{\alpha_s}{4\pi}\right)^2 C_i^{(2)} \, .
\end{eqnarray}
We find
\begin{eqnarray}
\label{eq:Cis}
C_1^{(0)}&=&1,\,\,\, C_2^{(0)}=C_3^{(0)}=0 \, , \\\nonumber
C_i^{(1)}&=&R_{(0),i}^{(1)}\, ,\,\,
C_i^{(2)}=R_{(0),i}^{(2)}+C_F^2K_{i,1}+C_Fn_hT_RK_{i,2} \,\, (i=1,2) \, , \\\nonumber
C_3^{(1)}&=&\frac{1-\hat{s}}{2}R_{(0),3}^{(1)},\,\,
C_3^{(2)}=\frac{1-\hat{s}}{2}\left(R_{(0),3}^{(2)}+
C_F^2K_{3,1}+C_Fn_hT_RK_{3,2}\right) \, ,
\end{eqnarray}
where the $R_{(0),j}^{(k)}$ were given in Section \ref{sec:TwoLoop}, 
and functions $K_{i,1}$  and  $K_{i,2}$  read
\begin{eqnarray}
\nonumber
K_{1,1}&=&-54-\frac{49 \pi ^2}{48}-\frac{\pi
   ^4}{160}-60 F_4-\frac{13 \pi ^2
   F_4}{12}+\frac{18
   F_4}{\hat{s}}+\frac{\pi ^2
   F_4}{12 \hat{s}}-\frac{31
   F_5}{2}-\frac{\pi ^2
   F_5}{6}+
      \\\nonumber &&
   \frac{13 F_5}{2
   \hat{s}}-
   8F_6+\frac{F_6}{\hat{s}}-2 F_7-63
   F_{10}-\pi ^2 F_{10}+\frac{29
   F_{10}}{\hat{s}}-22
   F_{11}+\frac{4 F_{11}}{\hat{s}}-
\\\nonumber &&
 8 F_{12}-28 F_{13}+
   \frac{6
   F_{13}}{\hat{s}}-8 F_{14}-12
   F_{15}-68 F_{17}+\frac{16
   F_{17}}{\hat{s}}-24 F_{18}-32
   F_{19}-
   \\\nonumber &&
   32 F_{20}-80
   F_{21}+\frac{5
   \zeta(3)}{3}+\frac{4}{3}
   F_4 \zeta(3)+
   \\\nonumber &&
   L \left(-78-\frac{5 \pi ^2}{4}-71
   F_4-\pi ^2 F_4+\frac{21
   F_4}{\hat{s}}-22 F_5+\frac{4
   F_5}{\hat{s}}-8 F_6-68
   F_{10}+\frac{16
   F_{10}}{\hat{s}}-
\right.\\  \nonumber && \left.   
   24 F_{11}-32
   F_{13}-80 F_{17}+\frac{4
   \zeta(3)}{3}\right)+
   \\  \nonumber &&
   L^2 \left(-\frac{97}{2}-\frac{\pi
   ^2}{2}-38 F_4+\frac{6
   F_4}{\hat{s}}-12 F_5-40
   F_{10}\right)+L^3 \left(-\frac{50}{3}-\frac{40
   F_4}{3}\right)-\frac{10 L^4}{3}\, ,
\end{eqnarray}
\begin{eqnarray}
\nonumber
K_{1,2}&=&-\frac{5 \pi ^2}{18}-\frac{2 \pi ^2
   F_4}{9}+\frac{4
  \zeta(3)}{9}+L \left(-16-\frac{2 \pi ^2}{3}-8
   F_4+\frac{8 F_4}{3
   \hat{s}}-\frac{16
   F_5}{3}-\frac{32
   F_{10}}{3}\right)-
   \\  \nonumber && 
   L^2 \left(20+16 F_4\right)-\frac{112 L^3}{9}\, ,
\end{eqnarray}
\begin{eqnarray}
\nonumber
K_{2,1}&=&\frac{18}{\hat{s}}+\frac{\pi ^2}{6
   \hat{s}}-\frac{9
   F_4}{\hat{s}^2}-\frac{\pi ^2
   F_4}{6 \hat{s}^2}+\frac{17
   F_4}{\hat{s}}+\frac{\pi ^2 F_4}{6
   \hat{s}}-\frac{7
   F_5}{\hat{s}^2}+\frac{7
   F_5}{\hat{s}}-\frac{2
   F_6}{\hat{s}^2}+\frac{2
   F_6}{\hat{s}}-\frac{22
   F_{10}}{\hat{s}^2}+
\\  \nonumber &&    
   \frac{22
   F_{10}}{\hat{s}}-\frac{8
   F_{11}}{\hat{s}^2}+\frac{8
   F_{11}}{\hat{s}}-\frac{12
   F_{13}}{\hat{s}^2}+\frac{12
   F_{13}}{\hat{s}}-\frac{32
   F_{17}}{\hat{s}^2}+\frac{32
   F_{17}}{\hat{s}} +
   \\  \nonumber &&   
   L \left(\frac{26}{\hat{s}}-\frac{18
   F_4}{\hat{s}^2}+\frac{26
   F_4}{\hat{s}}-\frac{8
   F_5}{\hat{s}^2}+\frac{8
   F_5}{\hat{s}}-\frac{32
   F_{10}}{\hat{s}^2}+\frac{32
   F_{10}}{\hat{s}}\right)+
   L^2
   \left(\frac{12}{\hat{s}}-\frac{12
   F_4}{\hat{s}^2}+\frac{12
   F_4}{\hat{s}}\right)\, ,
\end{eqnarray}
\begin{eqnarray}
\nonumber
K_{2,2}&=&L \left(\frac{16}{3
   \hat{s}}-\frac{16 F_4}{3
   \hat{s}^2}+\frac{16 F_4}{3
   \hat{s}}\right)\, ,
\end{eqnarray}
\begin{eqnarray}
\nonumber
K_{3,1}&=&-\frac{18}{\hat{s}}-\frac{\pi ^2}{6
   \hat{s}}+\frac{9
   F_4}{\hat{s}^2}+\frac{\pi ^2
   F_4}{6 \hat{s}^2}-\frac{53
   F_4}{\hat{s}}-\frac{\pi ^2 F_4}{3
   \hat{s}}+\frac{7
   F_5}{\hat{s}^2}-\frac{20
   F_5}{\hat{s}}+\frac{2
   F_6}{\hat{s}^2}-\frac{4
   F_6}{\hat{s}}+\frac{22
   F_{10}}{\hat{s}^2}-
\\  \nonumber &&      
   \frac{80
   F_{10}}{\hat{s}}+\frac{8
   F_{11}}{\hat{s}^2}-\frac{16
   F_{11}}{\hat{s}}+\frac{12
   F_{13}}{\hat{s}^2}-\frac{24
   F_{13}}{\hat{s}}+\frac{32
   F_{17}}{\hat{s}^2}-\frac{64
   F_{17}}{\hat{s}}+\\  \nonumber &&
   L \left(-\frac{26}{\hat{s}}+\frac{18
   F_4}{\hat{s}^2}-\frac{68
   F_4}{\hat{s}}+\frac{8
   F_5}{\hat{s}^2}-\frac{16
   F_5}{\hat{s}}+\frac{32
   F_{10}}{\hat{s}^2}-\frac{64
   F_{10}}{\hat{s}}\right)+
\\  \nonumber &&   
   L^2\left(-\frac{12}{\hat{s}}+\frac{1
   2 F_4}{\hat{s}^2}-\frac{24
   F_4}{\hat{s}}\right)\, ,
\end{eqnarray}
\begin{eqnarray}
\nonumber
K_{3,2}&=&   L \left(-\frac{16}{3
   \hat{s}}+\frac{16 F_4}{3
   \hat{s}^2}-\frac{32 F_4}{3
   \hat{s}}\right)\, .
\end{eqnarray}
The terms proportional to the explicit factors of $n_h$ 
in (\ref{eq:Cis}) stem from converting 
the results of the renormalized form factors $D_i$ from the
five-flavor to the four-flavor theory.  As a final check, we have 
confirmed that the $\mu$-dependence in the $C_i$ is such that 
the renormalization-group equation (\ref{eq:RGH}) is satisfied.


\section{Conclusions}
\label{sec:Conclusions}
We have presented results for the short-distance Wilson coefficients
needed to complete the calculation of partial decay rates 
in $\bar B \to X_u  \ell \bar \nu $
at NNLO in $\alpha_s$ and to leading order in $1/m_b$, for decay
kinematics limited to the shape-function
region.  The technical challenge was to compute the two-loop QCD
corrections to the semi-leptonic $b\to u$ transition current.  
To do this, we used the Laporta algorithm to perform a reduction
to master integrals, which were solved using the 
method of differential equations. We then performed a matching
calculation from QCD onto SCET to translate these results into 
the Wilson coefficients needed to compute the
hard function  in the factorization formula (\ref{eq:VagueFact}) at
NNLO.  In a companion paper, we shall perform an analysis of 
partial decay rates with arbitrary 
kinematic cuts at NNLO, and study the  implications on the 
determination of $|V_{ub}|$ from inclusive decays.

{\bf Acknowledgments}: We are grateful to Matthias Neubert for 
collaboration on the early stages of this work. 
The work of H.M. Asatrian was partially 
supported by ISTC A-1606 program.  C.G. is partially supported 
by the Swiss National Foundation as
well as EC-Contract MRTN-CT-2006-035482 (FLAVIAnet).
The Center for Research and Education in Fundamental Physics (Bern) is
supported by the ``Innovations- und Kooperationsprojekt C-13 of
the Schweizerische Universit\"atskonferenz SUK/CRUS''.

\vspace{0.4cm}
{\bf Note Added}: After our calculation was completed, the paper 
\cite{Bonciani:2008wf} appeared, where the UV-renormalized two-loop corrections
to the $b\to u$ current were presented.  We have compared with 
their results and found agreement with those given in Section
\ref{sec:TwoLoop}.


\begin{thebibliography}{99}


\bibitem{Korchemsky:1994jb}
  G.~P.~Korchemsky and G.~Sterman,
  Phys.\ Lett.\  B {\bf 340}, 96 (1994)
  [arXiv:hep-ph/9407344].

\bibitem{Akhoury:1995fp}
  R.~Akhoury and I.~Z.~Rothstein,
  Phys.\ Rev.\  D {\bf 54} (1996) 2349
  [arXiv:hep-ph/9512303].

\bibitem{Bauer:2003pi}
  C.~W.~Bauer and A.~V.~Manohar,
  Phys.\ Rev.\  D {\bf 70} (2004) 034024
  [arXiv:hep-ph/0312109].

\bibitem{Bosch:2004th}
  S.~W.~Bosch, B.~O.~Lange, M.~Neubert and G.~Paz,
  Nucl.\ Phys.\  B {\bf 699} (2004) 335
  [arXiv:hep-ph/0402094].

\bibitem{Lange:2005yw}
  B.~O.~Lange, M.~Neubert and G.~Paz,
  Phys.\ Rev.\  D {\bf 72} (2005) 073006
  [arXiv:hep-ph/0504071].


\bibitem{Bauer:2000ew}
 C.~W.~Bauer, S.~Fleming and M.~E.~Luke,
  Phys.\ Rev.\  D {\bf 63}, 014006 (2001)
  [arXiv:hep-ph/0005275].

\bibitem{Bauer:2000yr}
  C.~W.~Bauer, S.~Fleming, D.~Pirjol and I.~W.~Stewart,
  Phys.\ Rev.\  D {\bf 63}, 114020 (2001)
  [arXiv:hep-ph/0011336].


\bibitem{Beneke:2002ph}
  M.~Beneke, A.~P.~Chapovsky, M.~Diehl and T.~Feldmann,
  Nucl.\ Phys.\  B {\bf 643}, 431 (2002)
  [arXiv:hep-ph/0206152].

\bibitem{Neubert:1993ch}
  M.~Neubert,
  Phys.\ Rev.\  D {\bf 49}, 3392 (1994)
  [arXiv:hep-ph/9311325];  
  M.~Neubert,
  Phys.\ Rev.\  D {\bf 49}, 4623 (1994)
  [arXiv:hep-ph/9312311].

\bibitem{Bigi:1993ex}
  I.~I.~Y.~Bigi, M.~A.~Shifman, N.~G.~Uraltsev and A.~I.~Vainshtein,
  Int.\ J.\ Mod.\ Phys.\  A {\bf 9}, 2467 (1994)
  [arXiv:hep-ph/9312359].


\bibitem{Lee:2004ja}
  K.~S.~M.~Lee and I.~W.~Stewart,
  Nucl.\ Phys.\  B {\bf 721} (2005) 325
  [arXiv:hep-ph/0409045].

\bibitem{Bosch:2004cb}
  S.~W.~Bosch, M.~Neubert and G.~Paz,
  JHEP {\bf 0411}, 073 (2004)
  [arXiv:hep-ph/0409115].

\bibitem{Beneke:2004in}
  M.~Beneke, F.~Campanario, T.~Mannel and B.~D.~Pecjak,
  JHEP {\bf 0506} (2005) 071
  [arXiv:hep-ph/0411395].


\bibitem{Becher:2006qw}
  T.~Becher and M.~Neubert,
  Phys.\ Lett.\  B {\bf 637}, 251 (2006)
  [arXiv:hep-ph/0603140].



\bibitem{Grozin:1994ni}
  A.~G.~Grozin and G.~P.~Korchemsky,
  Phys.\ Rev.\  D {\bf 53} (1996) 1378
  [arXiv:hep-ph/9411323].


\bibitem{Becher:2005pd}
  T.~Becher and M.~Neubert,
  Phys.\ Lett.\  B {\bf 633}, 739 (2006)
  [arXiv:hep-ph/0512208].




\bibitem{Steinhauser:2002rq}
  M.~Steinhauser,
  Phys.\ Rept.\  {\bf 364}, 247 (2002)
  [arXiv:hep-ph/0201075].




\bibitem{Ali:2007sj}
  A.~Ali, B.~D.~Pecjak and C.~Greub,
  Eur.\ Phys.\ J.\  C {\bf 55}, 577 (2008)
  [arXiv:0709.4422 [hep-ph]].


\bibitem{Laporta:2001dd}
  S.~Laporta,
  Int.\ J.\ Mod.\ Phys.\  A {\bf 15} (2000) 5087
  [arXiv:hep-ph/0102033].

\bibitem{Tkachov:1981wb}
  F.~V.~Tkachov,
  Phys.\ Lett.\  B {\bf 100}, 65 (1981).

\bibitem{Chetyrkin:1981qh}
  K.~G.~Chetyrkin and F.~V.~Tkachov,
  Nucl.\ Phys.\  B {\bf 192}, 159 (1981).

\bibitem{Anastasiou:2004vj}
  C.~Anastasiou and A.~Lazopoulos,
  JHEP {\bf 0407}, 046 (2004)
  [arXiv:hep-ph/0404258].

\bibitem{Huber:2005yg}
  T.~Huber and D.~Maitre,
  Comput.\ Phys.\ Commun.\  {\bf 175} (2006) 122
  [arXiv:hep-ph/0507094].

\bibitem{Huber:2007dx}
  T.~Huber and D.~Maitre,
  Comput.\ Phys.\ Commun.\  {\bf 178}, 755 (2008)
  [arXiv:0708.2443 [hep-ph]].

\bibitem{Remiddi:1997ny}
  E.~Remiddi,
  Nuovo Cim.\  A {\bf 110} (1997) 1435
  [arXiv:hep-th/9711188].

\bibitem{Argeri:2007up}
  M.~Argeri and P.~Mastrolia,
  Int.\ J.\ Mod.\ Phys.\  A {\bf 22} (2007) 4375
  [arXiv:0707.4037 [hep-ph]].


\bibitem{Remiddi:1999ew}
  E.~Remiddi and J.~A.~M.~Vermaseren,
  Int.\ J.\ Mod.\ Phys.\  A {\bf 15} (2000) 725
  [arXiv:hep-ph/9905237].

\bibitem{Maitre:2005uu}
  D.~Maitre,
  Comput.\ Phys.\ Commun.\  {\bf 174}, 222 (2006)
  [arXiv:hep-ph/0507152].

\bibitem{Binoth:2003ak}
  T.~Binoth and G.~Heinrich,
  Nucl.\ Phys.\  B {\bf 680} (2004) 375
  [arXiv:hep-ph/0305234].

\bibitem{Bogner:2007cr}
  C.~Bogner and S.~Weinzierl,
  Comput.\ Phys.\ Commun.\  {\bf 178}, 596 (2008)
  [arXiv:0709.4092 [hep-ph]].


\bibitem{Bell:2006tz}
  G.~Bell,
  arXiv:0705.3133 [hep-ph].

\bibitem{Bell:2007tv}
  G.~Bell,
  Nucl.\ Phys.\  B {\bf 795} (2008) 1
  [arXiv:0705.3127 [hep-ph]].


\bibitem{Neubert:2004dd}
  M.~Neubert,
  Eur.\ Phys.\ J.\  C {\bf 40}, 165 (2005)
  [arXiv:hep-ph/0408179].


\bibitem{Korchemskaya:1992je}
  I.~A.~Korchemskaya and G.~P.~Korchemsky,
  Phys.\ Lett.\ B {\bf 287}, 169 (1992).

\bibitem{Bonciani:2008wf}
  R.~Bonciani and A.~Ferroglia,
  arXiv:0809.4687 [hep-ph].



\end{thebibliography}
\end{document}